\documentclass[11pt,twoside]{article}
\usepackage{graphicx,wrapfig,epsfig} 
\usepackage{amsmath}
\usepackage{amssymb}
\usepackage{cite}

\begin{document}
\newcommand{\pst}{\hspace*{1.5em}}

\newcommand{\be}{\begin{equation}}
\newcommand{\ee}{\end{equation}}
\newcommand{\bm}{\boldmath}
\newcommand{\ds}{\displaystyle}
\newcommand{\bea}{\begin{eqnarray}}
\newcommand{\eea}{\end{eqnarray}}
\newcommand{\ba}{\begin{array}}
\newcommand{\ea}{\end{array}}
\newcommand{\arcsinh}{\mathop{\rm arcsinh}\nolimits}
\newcommand{\arctanh}{\mathop{\rm arctanh}\nolimits}
\newcommand{\bc}{\begin{center}}
\newcommand{\ec}{\end{center}}

\thispagestyle{plain}

\label{sh}

\begin{center} {\Large \bf
\begin{tabular}{c}
A ``SPOOF LOOPHOLE" CONTRA NONLOCALITY
\end{tabular}
 } \end{center}

\bigskip

\bigskip

\begin{center} {\bf
A. F. Kracklauer$^{1*}$ 
}\end{center}

\medskip

\begin{center}
{\it
$^1$Bauhaus Universit\"at; Weimar, Germany

\smallskip
\today}
\smallskip

$^*$Corresponding author e-mail: af.kracklauer@web.de\\
\end{center}

\begin{abstract}\noindent
  A model for experiments testing Bell Inequalities is presented that does not
  involve nonlocal effects. It constitutes essentially a physical explanation
  of a \ ``loophole'' in the logic of these experiments, which, if not
  excluded, in principle nullifies the conventional conclusions, namely, that
  some kind of nonlocal interrelationship is intrinsic to quantum mechanics.
  The mechanism of the model ``spoofs'' the results predicted by conventional
  analysis employing quantum principles, without, however, using any of them.
\end{abstract}

\medskip


\section{Introduction}
\pst

Results from experiments testing the conclusion of what has become known as Bell's
Theorem are understood to verify that Nature (as described by Quantum Mechanics
--- QM) exploits nonlocal (and nonconventional) correlations. This conclusion,
which has been credited with being perhaps the most significant result of XX
century Physics{\cite{HS75}}, is as strong as the relevant experiments are
unambiguous. Nowadays, however, it is thought that these experiments are in fact
possibly restricted in their conclusiveness by several ``loopholes,'' which are
features of the experiments that {\emph{may}} invalidate the logical inevitability
of their interpretation (vice theory {\emph{per se}}). The loopholes considered so
far are not exposed by clues in the data, but rather, are aspects of the
experimental design, or natural processes which have not yet been taken into
account, and which, if present in fact, could lead to invalid conclusions, even
when the taken data is compatible with the patterns predicted by QM. \ Such
loopholes challenge the experimenter to find designs precluding these
weak structural features, so that conclusions cannot be put in doubt.

It is the purpose here to describe a new possible loophole of a slightly
different character in that it does lead to an otherwise spurious signature in
the data itself. \ The processes involved essentially spoof those correlation
patterns in the data which are predicted by QM and seem to support the current
common interpretation of tests of Bell inequalities verifying the existence of
nonlocality.

\section{The underlying physical cause}

The key physical feature in this new ``spoof-loophole'' pertains to the
source of correlated pairs of the signals. At present a very common source is
a nonlinear crystal in which a stimulus beam generates two output signals
with frequencies fixed by ``phase matching conditions'', and correlated  
polarization by cause of properties of the crystal (parametric down
conversion --- PDC){\cite{GRW,BEZ00}}. \ In general, it is considered
tacitly, that PDC crystals respond to stimulus beams as a single unit and
release one pair of output signals at a time --- at least at very low
intensities.  In other words, it is silently assumed, that by reducing the
intensity of the stimulus beam, noise, which is due to (partially) coincident
pairs, is reduced. 

The particular feature exploited herein, and the source of the ``spoof
loophole,'' is rejection of this particular silent assumption; namely, it is
taken here that there are multiple independent emission centers in such
crystals, each emitting a correlated pair of daughter signals at random,
possibly at overlapping, intervals, so that there will occur ``illegitimate''
coincidences, i.e., those not within one pair, but between disparate pairs. This
is equivalent to taking it, that a reduction in stimulus beam intensity is
linear in all consequences, so that it does not by itself tend to isolate
individual pairs, but just stretches the (overlapping) distribution of pairs
over longer time intervals. (This feature may have independent deleterious
consequences; see below.)

 Given this explicit assumption, it shall be shown, that filtering the data stream to
identify (presumably legitimate) coincidences by means of selecting an ever narrower
coincidence window actually preferentially selects illegitimate coincidences in very
close approximation to the ratio expected on the basis of calculations motivated by
QM.

\section{Quantum structure}

The essential quantum structure involved in this issue is captured by two
basic principles: Born's probabilistic interpretations of wave functions: $P
(x) = \psi^{^{_{\ast^{}}}} \psi$, i.e., the modulus of a wave function equals
a probability of presence; and the basic law of photo-detection: $I \propto
E^2$, i.e., a photo-current is proportional to the eliciting electric field
squared. Loosely speaking, for photons wave functions have been identified
with the electric field vector. 

This interpretation scheme remains self consistent so long as the wave functions to
which it is applied are ontologically acceptable ``as is," that is, so long as they can
be interpreted without need to call on the ``projection hypothesis.''  However, wave
functions used for the correlated pairs in experimental tests of Bell inequalities,
such as the singlet state 

\begin{equation}
  \psi = \frac{1}{\sqrt{2}} \Bigl(\langle v_l |\langle h_r | - \langle
  h_l | \langle v_r | \Bigr), 
\end{equation}
which is used to describe the signals emitted by a PDC, lead to inconsistencies
with the above principles for interpretation, as shall be argued below.  \ Here
$\left\langle v_l | \right.$ denotes the ``vertical'' channel on the ``left'',
whereas $\left\langle h_r | \right.$ the ``horizontal'' on the right, etc. Wave
functions of this character are essential, because they are rotationally
invariant, and this invariance property is an empirical fact (with caveat; see
below); but, this property also mandates the projection hypothesis to satisfy
the likewise empirical fact, that the existent (i.e., observed) constituent
single events cannot be an ambiguous combination of mutually exclusive states,
e.g., $\left\langle v_l | + \left\langle h_l| \right. \right.$.  To reconcile
this discrepancy, historically a ``projection postulate" has been introduced,
according to which it is asserted, that the act of measurement itself randomly
"projects" one of the sub-states to ``reality" and annihilates the other
substrate, thereby giving, for example either $\left\langle v_l | \right.$ or
$\left\langle h_l | \right.$. This matter, being irascibly counter intuitive,
has been at the focus of vast amounts of criticism and speculation.

The distressing feature here is seen immediately when Born's Principle is
applied to this wave function, Eq.~(1), to get:
\begin{equation}
  \psi^{_{} \ast} \psi = \frac{1}{2}\Bigl(\left\langle v_l | v_l \right\rangle \left\langle h_r
  | h_r \right\rangle  -2 \left\langle v_l | h_{_r} \right\rangle
  \left\langle h_{_l} | v_r \right\rangle + \left\langle v_r | v_r
  \right\rangle \left\langle h_l | h_l \right\rangle \Bigr).
\end{equation}
The first and the third terms can be interpreted as the product of the squares
of two wave functions, and as such, can represent a joint probability of
coincident events, therefore they are consistent with the Born Principle in that
they yield admissible coincidence probabilities. Moreover, the first and third
individual terms fit within the pattern of the photo-detection law as each term
specifies the intensity of a joint photo-current in terms of the square of 
electric fields.

The middle term, however, as the product of four different wave functions (or
four distinct electric fields), neither can be seen as a probability according to Born's
Principle, nor does it conform to the photo-detection law. Nevertheless, it is
mathematically essential to procure a rotational invariant state.  It also
seems to encapsulate much that is mysterious in Quantum Mechanics. The model
proposed here rationalizes this term.

\section{The loophole mechanism}

For the model we assume the following: Each signal pair comprises two
distinctly directed electromagnetic pulses of equal duration and fully
(anti)correlated polarization states. In turn, each pulse, said schematically
to be sent to the right or left, is taken to elicit a single photo-electron in
a detector (subsequently amplified, etc.) at a random time, $t_0$, occurring
with a uniform random distribution over the pulse length $l$. That is, we
assume the usual character of photo-electron elicitation. Further, it is taken also,
that a signal pair has a random offset relative to any other pair. 

Now, each photo-electron (hit) pair generated by a particular signal pair shall be
denoted as ``legitimate''; whereas pairs of hits from disparate, but overlapping
signal pairs, as ``illegitimate'' coincidences. If observation is limited to
within a window, $w_0$, shorter than the pulse length, $l$,  it may well happen, that
one of the hits from a particular pair is inside, and the other hit from that same
pair is outside of the window, and that the nearest time coincidence within the
window occurs (accidentally) between hits from two disparate pairs
(illegitimately) while the legitimate coincidence remains unseen or excluded from  the
analysis.

Illegitimate coincidences are crucial within the spoof model because some of them can
be associated with the middle term in Eq.~(2). \ This can be understood as
follows. Polarizers oriented at a finite angle to the polarization vector of the
input signal, according to Malus' Law, divide output between two channels which
can be denoted: \ ``vertical'' and ``horizontal.'' \ ``Vertical'' is taken with
respect to the polarizer axis, which itself is at an angle \ $\theta_i$ \ with respect to the
laboratory, and therefore the source crystal's, or the input signal's, vertical. In
turn, according to Malus' Law, the intensities directed into photo-detectors
reading the outputs of the polarizer are proportional to \  $\cos^2 (\theta)$
in the vertical output channel and $\sin^2(\theta)$ in the horizontal channel.

When the intensity is so low, that the output signal can elicit only one
photo-electron in a detector, the explanation based on quantum theory is that a
photon has been directed into one of these two channels with probability given by
$\cos^2 \theta$ or $\sin^2 \theta$. \ Either way, the intensity of the output
signal in the two channels is scaled by these (Malus) factors. This implies, that
for any single photo-electron actually detected in either channel, there are two
possible signals impinging on the polarizer that might have been its progenitor,
e.g., a vertical signal from the source projected into the vertical channel of the
polarizer with  with probability proportional to $\cos^2 \theta$, or a horizontal
signal from the source crystal projected into the vertical channel with probability
proportional to $\sin^2 \theta$.  A corresponding situation holds for the
horizontal channel of the polarizer with exchanged Malus factors.

This ambiguity in the true source of the signal emerging from polarizers
(PBS's) allows one, on the basis of the identities:
\begin{equation}
  \cos (\theta) = \sin (\theta + \pi / 2),\qquad {and} \qquad \sin (\theta) = -
  \cos (\theta + \pi / 2),
\end{equation}
to write
\begin{equation}
  \left\langle v_i \left| \equiv \left\langle h_j \left|, \qquad{and}\qquad \left\langle
  h_i \left| \equiv - \left\langle v_j | . \right. \right. \right. \right. \right.
  \right. \right.
\end{equation}

Thus, when these symbolic associations are made for the middle terms of  Eq.
(2), there arises a {\it formal} identification between (skew) illegitimate
coincidences and the mixed or cross terms. The physics supporting this idea is
clear and credible; real electric fields from real signals actually fall on the
detectors and elicit photo-electrons, which accounts for the additional
coincidences (relative to the number of legitimate coincidences) that are indeed
counted in experiments. The pure symbol manipulation represented by Eqs.~(3) and
(4) then permits associating (symbolically) these coincidences with those
``cross" terms in Eq.~(2). In other words, completely distinct events
(illegitimate coincidences) are attributed symbolically to the legitimate
events, so that this symbol combination fits into the structure of the Quantum
Mechanics (Born's `law') although they have a purely non quantum physical cause
in reality. 

All this is what  renders this expression rotationally invariant and compatible with
Born's interpretation and the photo-detection law. \ \ Moreover, this effect actually
explains classically the origin of physical rotational invariance.

Given the above, the next question is: does the ratio of proper to
improper illegitimate coincidences conform with the ratio satisfying Eq.~(2);
which is 1::1? 

There are two cases to consider: illegitimate pairs and legitimate pairs.

The probability density of illegitimate coincidences is just the product of
their individual, absolute probability densities, which can be calculated from
the average frequency of pair generation, $f$. First note that $f$ is the
numerical inverse of the time interval for which the expectation of elicitation
of one photo-electron in a detector (a ``hit'') equals $1$. Because the search
for the first hit is not confined to a window, its probability density is $f^{-
1}$. The (accidental) partner hits are uncorrelated with respect to the first
seen hit, which is the instant at which the window is, so to speak, opened. \
There is no {\emph{a priori}} reason that the frequencies, $f_n$, for distinct
signal pairs (different $n$), arising presumably in different locations within
the source crystal under various geometrical and electrodynamic conditions, must
be absolutely identical. \ Thus the total probability of one or the other illegitimate
alternative pairing is just the sum of the probability densities of the
possibilities integrated over the window, symbolically
\begin{equation}
  \int_0^{w_0} f^{- 1} \left( \sum_n f_n^{- 1} \right) d t = k w_0,
\end{equation}
where the sum is over all the possible (overlapping) disparate pairs. Note that,
in our mind's eye, the first hit was imagined herein as occurring on the left,
and then a coincidental hit on the right was sought; but, as we imagine
absolutely coincident pairs to be extremely improbable, it makes no difference
in which channel (side) the first hit is found.  As the data stream is analyzed
in sequence, if a hit has occurred on one side, then its partner is expected on
the other. Also note, there are generally two possible polarization states on
each side making a total of four combinations corresponding to the the four
terms of Eq.~(2).  

A calculation for legitimate coincidences is more complex, because the events
are not statistically independent, and a joint probability density is no
longer just the product of two absolute probability densities. \ The essential
question now is: what is the average likelihood of seeing a legitimate partner
hit, given that the first hit has occurred at $t_0$, if observation is
restricted to within the window of width ``$w_0$'' opened at $t_0$? This probability involves a
{\it conditional probability}, because the second pulse is correlated with the
first pulse.

Given that the hit which opened the window occurred on the left, say,
at $t_{0}$, it follows that its legitimate partner must be found
on the right within the remaining pulse length, that is within the
time interval $l-t_{0}$. Thus, the conditional probability for this
partner hit, is $1/(l-t_{0})$. Now, given this probability density
and initial hit, the accumulated probability within the window, $w_0$,
will equal $w_0/(l-t_{0})$ for every occurrence having this initial
instant for the first hit.

However, when the window width $w_0$ exceeds $l-t_{0}$, then this
condition probability must be $1$, which means that this conditional
probability comprises two segments:\begin{equation}
\int_0^{w_0}\rho(t_{0},w)dw=\rho(t_0|w_0)=\begin{cases}
\frac{w_0}{l-t_{0}},\quad & 0<t_{0}<l-w_0\\
1 & l-w_0\geq t_{0}\geq l\end{cases}\end{equation}

This is a two-dimensional density, dependent on the instant of the first hit
and the then opened window width, $w_0$, and which is variable over a triangular
region in which the window width is insufficient to cover the remaining pulse
length. We are most interested in the variation of the cumulative probability
of this expression as a function of the window width, $w$. \ It is the integral
of $\rho(t_{0}|w_0)$ over all times, $t_{0}$ of the first hit from $0$ to $l-t_0$,
or to the point at which the accumulated probability equals one, i.e., all
partner hits have been found: Now, it turns out, that for any given fixed
window width, $w_0$, all coincidences will have been registered when
$l-t_0=w_0$.  Thus, for computing the accumulated probability of seeing a
coincidence as a function of $w_0$ can be obtained by the integral:
\begin{equation}
 \int_{0}^{l-t_0}\rho(t_{0}|w_0)\,dt_{0}=\int_{0}^{w_0}\rho(t_{0}|w_0)\,dt_{0}
=-w_0\ln(l-w_0).
\end{equation} 
This is the accumulated probability of encountering legitimate
partners by filtering a data stream by searching in a window of width $w_0$ in
the partner channel opened at the instant when a hit is seen in either channel.

Here we observe, that irrespective of the density of illegitimate pairs (given there
is at least some) the ratio of their various types of coincidences corresponding to
the terms in Eq.~(2) is satisfied perfectly just by ordinary (but here for
illegitimate events) probabilities.  It is the {\it legitimate} coincidences which
spoil the proportions!  This in turn implies, that any procedure which preferentially
filters out these legitimate coincidences will cause the statistics to converge on
exactly those proportions seeming to (erroneously) validate  Eq.~(2).\vfill\eject

\begin{wrapfigure}{r}{85mm}{\includegraphics[width=76mm]{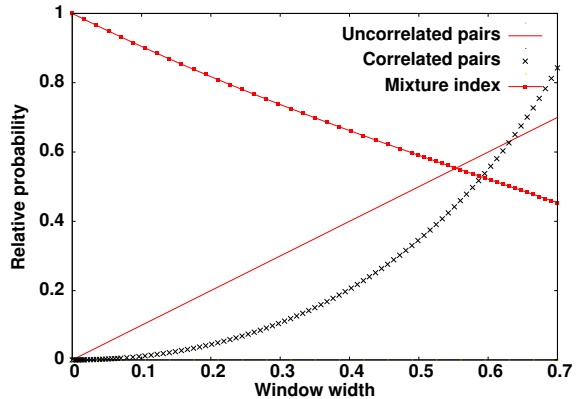}}
\caption{\small Comparison of the probabilities of legitimate with illegitimate
coincidences. Clearly, by reducing the window width, $w_0$ (in units of pulse
length $l$), legitimate coincidences are filtered out preferentially.  In
addition, the ``mixture index" giving the percentage of legitimate coincidences
in the total data set is shown.  This index reflects the effect of the spoof} 
\end{wrapfigure}

The relative effects of reducing the window width are illustrated
quantitatively in the figure.  The quantity of illegitimate coincidences from
uncorrelated pairs diminishes linearly with $w_0$, (where we consider the
weakest case, i.e., $k=1$). On the other hand, the number of legitimate
coincidences from correlated pairs diminishes more rapidly with {\it
decreasing} $w_0$.   As a consequence, reducing $w_0$ with the aim of purging
illegitimate coincidences, actually achieves just the opposite.

\begin{sloppypar}

The fact that the illegitimate coincidences cannot be totally removed from
the data set with finite $w_0$ has the consequence that feasible data sets
will always contain an admixture of legitimate coincidences, which, as
argued above, spoil rotational invariance.  The mixture index, namely:
$k/(k-\ln(1-w_0))$, on the graphic gives the percentage of the admixture;
and, as such, quantitatively reflects the degree to which data from
EPR/Bell inequality tests do not meet expectations from  Bell's. The two
most salient features of this sort are: 1) the failure to obtain the full
limit of $2\sqrt{2}$ as predicted by Bell's analysis, and 2) a residue of
rotational variance.  Both features have been reported; and, the shape of
the mixture index curve faithfully represents their variation as a function
of $w_0$, essentially constituting a signature within the data of the
presence of mechanisms of the ``spoof loophole."  \cite{TW07, AK,
AFK}\end{sloppypar}

\section{Discussion and conclusions}

\

We note that this, at first glance counter intuitive, statistical effect may not
be the only contributor to undermining the customary (Bell's) conclusion. If
there is any systematic difference in the optical path lengths of the two arms,
the consequence is the same: narrowing the window preferentially
\emph{excludes} legitimate coincidences. Given that the exact timing of
emission of a photo-electron within the stimulating pulse length is a random
variable, it becomes very difficult to determine the underlying pulse head
arrival times, which could serve as calibration for fixing the optical path
lengths. In other words, it may well be impossible in principle to obtain
sufficiently identical optical path lengths.

In addition, we note, that our fundamental physical assumption introduces sever
practical challenges also.  Currently available detectors recover insufficiently
fast after a detection to permit excluding effects sullying the statistics of
the data.\cite{KOG}  In addition, it has been shown, that the arcanum of
detector thresholds can be arranged hypothetically so as to exceed the Bell
locality limit.\cite{GA} These effects are serious impediments to absolutely
conclusive deductions from EPR-Bell inequality texts.  Although, given the
"extraordinary" character of the claim for nonlocality, it would be prudent to
require equally extraordinary proof, sociologically it seems that simple
intuitively unlikely possible causes for loopholes are given only ``academic"
credibility, so that ``for all practical purposes" nonlocality is accepted
nowadays as a reality.  

Nevertheless, it can be argued reasonably, especially given its signature in the
data, that unless the mechanism of the spoof loophole can be precluded {\emph{a
priori}}, experiments testing EPR phenomena or Bell inequalities are
inconclusive;  the existence of nonlocality has not been empirically
substantiated and, therefore, is not yet beyond question, even just ``for
practical purposes."

\end{document}